\def\zs{S}
\def\rs{R}
\def\rdr{\mu}
\def\diffr{{\Delta r}}
\def\vdiffr{{\bf \Delta r}}
\def\Gphi{{G_\phi}}
\def\Gvel{{\bf G_v}}
\def\Gdel{{G_\delta}}
\def\etal{{\frenchspacing\it et al.}}
\def\ie{{\frenchspacing\it i.e.}}

\def\etc{{\frenchspacing\it etc.}}
\def\vzero{{\bf 0}}
\def\vv{{\bf v}} \def\vr{{\bf r}} \def\vk{{\bf k}} 
\def\vs{{\bf s}}
\def\rhat{\widehat{\vr}}

\def\zhat{\widehat{\bf z}}

\def\l{\ell}

\def\DU{\Delta U}

\def\ehat{\widehat{e}}
\def\d{\delta}
\def\dd{\delta}

\def\expec#1{\langle#1\rangle}

\def\suml{\sum_{\l=0}^{\infty}}

\def\vr{{\bf r}}

\def\alph{\alpha}
\def\gamm{\gamma}
\def\al{a_{\l}}
\def\Pl{P_{\l}}
\def\dr{\partial_r}
\def\dx{\partial_x}
\def\dy{\partial_y}
\def\dz{\partial_z}
\def\A{A}
\def\kms{\,{\rm km/s}}

\def\notV{V\hskip-0.22cm \backslash}

\def\spose#1{\hbox to 0pt{#1\hss}}
\def\simlt{\mathrel{\spose{\lower 3pt\hbox{$\mathchar"218$}}
     \raise 2.0pt\hbox{$\mathchar"13C$}}}
\def\simgt{\mathrel{\spose{\lower 3pt\hbox{$\mathchar"218$}}
     \raise 2.0pt\hbox{$\mathchar"13E$}}}
\def\simpropto{\mathrel{\spose{\lower 3pt\hbox{$\mathchar"218$}}
     \raise 2.0pt\hbox{$\propto$}}}

\def\addr#1{{\small\it #1}}
\def\auth#1{{#1}}

\documentstyle[11pt,epsf]{article}
\def\beq#1{\begin{equation}\label{#1}}
\def\eeq{\end{equation}}
\def\beqa#1{\begin{eqnarray}\label{#1}}
\def\eeqa{\end{eqnarray}}
\def\eq#1{equation~(\ref{#1})}
\def\Eq#1{Equation~(\ref{#1})}

\def\eqnum#1{~(\ref{#1})}

\def\deltar{\delta_{\rs}}
\def\ds{\delta_{\zs}}
\def\nbar{{\bar n}}
\def\Wh{\widehat{W}}
\def\Gdelh{\widehat{\Gdel}}
\def\Mpc{{\rm Mpc}}

\begin{document}

\begin{titlepage}   

\hfill MPI-PhT/94-61

\begin{center}

\vskip0.9truecm
{\bf

REAL-SPACE COSMIC FIELDS FROM REDSHIFT-SPACE DISTRIBUTIONS:

A GREEN FUNCTION APPROACH%
\footnote{Published in {\it ApJ}, {\bf 453}, 533 (November 10, 1995).
Submitted September 20, 1994.\\
Available from
{\it h t t p://www.sns.ias.edu/$\tilde{~}$max/zspace.html} 
(faster from the US) and from\\
{\it h t t p://www.mpa-garching.mpg.de/$\tilde{~}$max/zspace.html} 
(faster from Europe).\\
}
}

\vskip 0.5truecm
  \auth{Max Tegmark}

  \smallskip
  \addr{Max-Planck-Institut f\"ur Physik, F\"ohringer
Ring 6, D-80805 M\"unchen}

  \addr{max@mppmu.mpg.de}
  \smallskip
  \vskip 0.2truecm

  \auth{B.\ C.\ Bromley}

  \smallskip
  \addr{Theoretical Astrophysics, MS B288, Los Alamos 
National Laboratory,}

  \addr{Los Alamos, NM 87545}

  \addr{bromley@eagle.lanl.gov}

  \smallskip

\end{center}


\def\taut{\tau}

\abstract{

We present a new method for reconstructing the cosmological 
density, peculiar velocity and peculiar gravitational 
potential on large scales from redshift data.
We remove the distorting effects of 
line-of-sight peculiar motions by using the linear theory
of gravitational instability, in which the potential is the 
solution to a linear partial differential equation first 
derived by Kaiser. We solve this equation by deriving its 
Green function; the Green functions for the peculiar velocity and
the real-space density follow directly.
Reconstruction of the cosmic fields is thus reduced to 
integration
over redshift space with the appropriate Green function 
kernel.
Our algorithm has a single input parameter,
$\beta \equiv \Omega^{0.6}/b$,
where $b$ is the linear bias factor and $\Omega$ is the 
cosmological density
parameter. 
We discuss 
the virtues of this method for error control, for 
estimating $\beta$, and for
constraining the bias mechanism.
}

\end{titlepage}

\section{Introduction}

Redshift surveys are testimony to the richness of structure 
in the universe on large scales (see, for instance, 
Giovanelli \& Haynes 1991).
The observed mass distribution in ``redshift space", 
the space defined by two sky coordinates and a radial 
recession velocity,
depends on both the real-space mass density and the peculiar 
velocity field, each of which are of great importance in 
cosmology.
The purpose of this paper is to present a new method for 
evaluating
both the density and velocity fields, as well as the peculiar 
gravitational
potential, directly from redshift data.

While quantifying the redshift-space distortions is difficult 
in general,
Kaiser (1987) has solved the problem for large scales 
($> 10 \, {\rm Mpc}$) in the linear regime of gravitational 
instability theory (Peebles 1993, \S 5):
The effect of linear peculiar flows on the 
observed redshift-space density field
is, loosely speaking,
to enhance the amplitude of fluctuation modes along the 
observer's line of sight by a factor of 
\beq{beta}
\beta = {\Omega^{0.6}}/{b} \, ,
\eeq
where the parameter $b$ admits the possibility that there is
a linear bias in the fluctuations of luminous matter relative
to the total mass distribution.
This dependence on $\Omega$ has been exploited
to estimate the cosmic mass density from redshift data 
(Hamilton 1992, 1993; 
Fisher, Scharf \& Lahav 1994; Cole, Fisher \& Weinberg 1994;
Bromley 1994).
For the purposes of removing redshift space distortions,
we will assume that $\beta$ is known or that it can be
estimated in some self-consistent way from the data.

The problem of reconstructing cosmic fields from 
redshift data has attracted considerable interest in the
last few years. 
Yahil et al.\ (1991), 
Taylor \& Rowan-Robinson (1993)
and
Gramann, Cen \& Gott (1994) 
have implemented iterative algorithms for real-space density 
and peculiar
velocity reconstructions in the linear or quasi-linear regimes.
In rough terms,
the iterative strategy is to assume that the redshift data 
approximate the 
real-space density, calculate the peculiar flows, then 
correct
the density accordingly. The procedure is repeated until
the real-space distribution and the velocity field converge 
to a
solution that is consistent with the observed redshifts.
A different strategy has been to expand all fields in spherical
harmonics, and then estimate each multipole individually
(Fisher, Scharf \& Lahav 1994; Nusser \& Davis 1994). 

Here
we start with Kaiser's equation for
redshift space distortions and solve for the Green functions 
that give the real-space density, the peculiar velocity 
field, and
the peculiar gravitational potential.
Thus, the reconstruction algorithm simply involves
integration over the redshift-space density
with
a Green function kernel.
The method is designed to hold on scales in the
linear regime under the assumption 
that light traces mass or that luminous matter is a linearly 
biased
sample of the total density field.
The technique is valid for any survey geometry.
In particular, we do not make Kaiser's ``plane-parallel'' 
approximation 
wherein the redshift survey is assumed to lie in a compact 
region at a large
distance from the observer (see Zaroubi \& Hoffman 1994).

The Green function method has distinct advantages over
previous techniques: it is a noniterative, linear
algorithm applicable to any survey geometry, and the work is 
done entirely in 
real space, not in the Fourier domain or with spherical 
harmonic coefficients.
Hence the
error bars due to edge effects can be 
calculated
in a straightforward and intuitive way.

This paper
is organized as follows.
After reviewing Kaiser's result and 
establishing some notation in Section 2, 
we derive the Green function for the
peculiar gravitational potential in Section 3, 
from which the kernels for
peculiar velocity and real-space density follow directly.
In Section 4, we specialize to the simple case
known as the ``distant survey approximation".
In Section 5, we discuss how the technique can be applied
to real-world situations where one faces
nonlinear clustering, shot noise and finite volume.
In Section 6, we give a numerical example of density reconstruction
from artificial redshift-space data. We conclude by 
summarizing possible
applications of the method for extracting $\Omega$ and 
constraining
the biasing mechanism.

\section{The partial differential equation}

The first step is to obtain an expression, analogous to 
a Poisson equation, which relates the real-space peculiar
gravitational potential $\phi$ to the
density fluctuation field $\d_\zs$ in redshift space
(we denote redshift-space quantities with
a subscript-$\zs$ while subscript-$\rs$ indicates real 
space).
This was first done by Kaiser (1987), but we give a brief 
derivation here for completeness.

The fluctuation field is defined in the usual way,
\beq{deldefEq}
\d = \left[ \rho(\vr) - \overline{\rho} \right] / 
\overline{\rho} \, ,
\eeq
where $\rho$ is the mass density and $\overline{\rho}$ is its 
mean.
Here, the density fields are associated with luminous matter, 
and we assume that
fluctuations in light are proportional to fluctuations in 
total mass
on the scales of interest. We use time units such that 
the Hubble constant $H_0=1$. 

Denoting the radial part of the peculiar velocity field $\vv$
at a real-space position $\vr$ by
\beq{UdefEq}
U(\vr) \equiv \vv(\vr)\cdot\rhat \, ,
\eeq
and the radial peculiar velocity in the observer's frame by%
\beq{DUdefEq}
\DU(\vr) \equiv [\vv(\vr) - \vv(\vzero)]\cdot\rhat \, ,
\eeq
the apparent position $\vs$ of a
galaxy in redshift space is
\beq{sEq}
\vs(\vr) \equiv \vr + \DU(\vr)\rhat \, .
\eeq
As long as the perturbations are sufficiently small
that this mapping is one-to-one,
the overdensities in redshift space and in real 
space are related by
\beq{DensityEq}
(1+\delta_\rs)d^3r = (1+\delta_\zs)d^3s \, .
\eeq
Evaluating the Jacobian of the mapping\eqnum{sEq} yields
\beq{JacobianEq}
d^3s = 
\left(1+{\DU\over r}\right)^2
(1+\dr U) d^3r \, ,
\eeq
where $\dr$ denotes the radial derivative $\partial/\partial 
r$.
Thus, to first order in $\vv$ and $\delta$, and
in the absence of selection effects
and survey limits, the equation
relating the densities in redshift space and in real space is
\beq{FullKaiserEq}
\delta_\zs = \delta_\rs - \dr U 
- 2 {U\over r} + 2 {\vv(0)\cdot\rhat\over r}\, .
\eeq
We can express the first three terms on the right hand side 
in terms of the peculiar gravitational potential, $\phi$.
The density $\delta_\rs$ and $\phi$ are related through the 
Poisson equation, 
\beq{PoissonEq}
\nabla^2 \phi = \A\delta_\rs,
\eeq
where (cf. Peebles 1993, eq. [5.107]) we have introduced the 
constant 
\beq{AEq}
\A = 4\pi G\rho_0 = {3\over 2}\Omega.
\eeq
In linear theory, the peculiar velocity is simply 
proportional to the gravitational force field, and is given 
by (cf. Peebles 1993, eq. [5.117])
\beq{vEq}
\vv = -{\beta\over\A}\nabla\phi.
\eeq
Thus the velocity terms are
\beq{dUeq}
\dr U = -{\beta\over A} \dr^2\phi \, ; \ \ \ \ \
{U\over r} = -{\beta \over \A r} \dr\phi \, .
\eeq
Substituting this into \eq{FullKaiserEq} yields what we will 
refer to as the {\it Kaiser equation},
\beq{SimplePDE}
D\phi = \A\delta'_\zs\, ,
\eeq
where the differential operator $D$ is defined as
\beq{DdefEq}
D \equiv \nabla^2 + \beta\dr^2 + 2\beta{1\over r}\dr\, .
\eeq
Here we have eliminated the last term in 
\eq{FullKaiserEq} by defining 
$\delta'_{\zs}$ as the redshift space density 
observed {in the cosmic rest frame}, the frame relative to 
which $\vv$ is defined. From 
the observed dipole anisotropy in the microwave 
background radiation, we know that the peculiar velocity of 
our solar system is (Smoot {\etal} 1992)
\beq{vzeroEq}
\vv(0) = 370 \pm 10 \,\kms
\eeq
in the direction $\alpha = 11.2^h$, $\delta=-7^{\circ}$. 
The cosmic variance in this dipole is merely a few km/s,
{\ie}, considerably smaller than the errors quoted above.
$\delta'_{\zs}$ can be directly 
computed from the redshift-space density we observe, 
$\delta_{\zs}$, either by 
using the first order correction derived above, 
\beq{deltasCorrEqOne}
\delta'_{\zs}(\vr) = 
\delta_{\zs}(\vr) - 2{\vv(0)\cdot\rhat\over r},
\eeq
or by simply adjusting the observed redshifts 
(radial galaxy positions):
\beq{deltasCorrEqTwo}
s' = s + \vv(0)\cdot\rhat.
\eeq
In what follows, we will drop the prime and let 
$\delta_{\zs}$ denote the corrected field.
It should be noted that this distinction is of only marginal importance. 
The last two terms in 
\eq{FullKaiserEq}, 
are usually negligible, and have been dropped by most authors.
We leave them in because they preserve the self-adjointness of the
operator $D$. It turns out that the approximation of
dropping them gives a more complicated solution, not a simpler one.

\section{The Green functions}

We now solve the Kaiser \eq{SimplePDE}.
Comparing the differential operator $D$ with the expression for
$\nabla^2$ in spherical coordinates shows that it is
simply the Laplacian 
with the radial part amplified by a factor 
$(1+\beta)$.
It is self-adjoint for all $\beta$, 
which means that 
its eigenfunctions are orthogonal, and that expanding the
fields in these functions (spherical Bessel functions
of non-integer order) provides a straightforward way of 
solving the equation. A related formal solution is given 
by Nusser \& Davis (1994), but for a slightly different 
operator yielding the solution in redshift space, not real space. 
Here we will instead give a more direct solution,
in real space, which renders orthogonal expansions unnecessary.

By the linearity of the problem, 
we can always write down a formal solution as
\beq{GdefEq}
\phi(\vr) = \A \int \Gphi(\vr,\vr') \d_\zs(\vr') d^3r' \, ,
\eeq
where the Green function (integral kernel) $\Gphi$ satisfies
\beq{Geq}
(D\Gphi)(\vr,\vr') = \dd^3(\vr'-\vr) \, 
\eeq
and the 
Dirac delta function 
is distinguished from the
fluctuation fields $\delta_\rs$ and $\delta_\zs$
by the absence of a subscript.
The constant $\A$ in Eq.\eqnum{GdefEq} was inserted to allow
us to think of $\Gphi$ as the operator inverse of $D$. 
Although $\Gphi$ is a function of six variables, there are in
fact only two quantities upon 
which the dependence is non-trivial.
Since the Kaiser equation is invariant under 
rotation and rescaling, it is easy to see that 
we can write 
\beq{SymmetryEq}
\Gphi(\vr,\vr') = {g\left(\rdr,r/r'\right)\over \sqrt{rr'}}
\eeq
for some function $g$, with the angle cosine $\rdr \equiv 
\rhat\cdot\rhat'$.
Let us utilize this property by
expanding $\Gphi$ in Legendre polynomials as 
\beq{LegendreExpansionEq}
\Gphi(\vr,\vr') = {1\over\sqrt{r r'}} 
 \suml(2\l+1)\Pl(\rdr)\al(r/r').
\eeq
The Kaiser equation implies that the radial
functions $\al(x)$ satisfy
\beq{alEq}
\al'' + {1\over x}
\al'- {1\over x^2}\left[{1\over 4}+{l(l+1)\over 1+\beta}\right]
\al = 
{\delta(x-1)\over 4\pi(1+\beta)}.
\eeq
Requiring $\al$ to be continuous and imposing the 
boundary conditions that $\al(x)$ remain finite
both as $x\to 0$ and as $x\to\infty$, this equation has the 
solution
\beq{alSolEq}
\al(x) = -{x^{\pm\alph_\l}\over
8\pi(1+\beta)\alph_\l},
\eeq
where the exponent is $+\alph_\l$ if $x\leq 1$ and 
$-\alph_\l$ if $x\geq 1$, with
\beq{alphaEq}
\alph_\l = {1\over 2}\sqrt{1 + 
{4\l(\l+1)\over 1+\beta}}\,.
\eeq
Substitution back into \eq{LegendreExpansionEq} gives
the solution
\beq{GsolEq}
\Gphi(\vr,\vr') = -{1\over 8\pi(1+\beta) \sqrt{rr'}} 
\suml
\left({2\l+1\over\alph_\l}\right)
\Pl(\rdr)\left({r_<\over r_>}\right)^{\alph_\l},
\eeq
where
$r_<$ and $r_>$ denote the smaller and larger of the
numbers $r$ and $r'$, respectively.
If we let $\beta\to 0$, the Kaiser equation approaches the
Poisson equation, 
$(2\l+1)/\alph_\l\to 2$ and
\eq{GsolEq} reduces to the familiar
\beq{GsolLimitEq}
\Gphi(\vr,\vr') = -{1\over 4\pi|\vr'-\vr|}.
\eeq

We now consider the peculiar velocity field $\vv$, which 
according to \eq{vEq} is proportional to the gradient of the 
peculiar gravitational potential $\phi$. 
Rather than computing $\nabla\phi$ numerically, it is 
convenient to define a Green function $\Gvel$ expressly for 
the
velocity field which satisfies
\beq{GvdefEq}
\vv(\vr) = \int \Gvel(\vr,\vr') \delta_\zs(\vr') d^3r' \,,
\eeq
where, evidently, 
\beq{vGeq}
\Gvel = \beta\nabla\Gphi \, .
\eeq
Taking the gradient of \eq{GsolEq} with respect to $\vr$ 
gives
\beqa{GGeq}
\Gvel(\vr,\vr') &=& -{\beta\over 8\pi(1+\beta)r'^2} \suml
\left({2\l+1\over\alph_\l}\right)
\left({r\over r'}\right)^{\gamm_\l-1} 
\\ \nonumber 
&\times&
\left\{\gamm_\l\Pl(\rdr)\ehat_r+\l\left[\rdr\Pl(\rdr)-P_{\l-
1}(\rdr)
\right]\ehat_\theta
\right\},
\eeqa
where $\gamm_\l\equiv -1/2 + \alph_l$ if $r<r'$ and
$\gamm_\l\equiv -1/2 - \alph_l$ if $r>r'$.
The coordinate
$\theta$ refers to a spherical coordinate system with the 
zenith in the $\vr'$-direction, {\ie}, where 
$\rhat\cdot\rhat' = \cos\theta$.

For completeness, we finally 
construct the Green function for the 
real-space density, $\Gdel$, such that
\beq{KdefEq}
\delta_\rs(\vr) = \int \Gdel(\vr,\vr') \delta_\zs(\vr') d^3r' 
\, .
\eeq
Since the Poisson \eq{PoissonEq} gives
\beq{KGeq}
\Gdel = \nabla^2 \Gphi \, ,
\eeq
taking the Laplacian of \eq{GsolEq} yields
\beqa{KsolEq}
\Gdel(\vr,\vr') &=& 
%
%
{1\over 8\pi(1+\beta) r'^3} \suml
\left({2\l+1\over\alph_\l}\right)
[\gamm(\gamm+1)-\l(\l+1)]
\Pl(\rdr)\left({r\over r'}\right)^{\gamm-2}\nonumber\\
%
%
&+&\left({3+2\beta\over 3+3\beta}\right)
\delta^3(\vr'-\vr)\,,
\eeqa
where the delta-function term represents the full 
contribution
of $\Gdel$ at $\vr = \vr'$.
As expected, 
\beq{KsolLimitEq}
\Gdel(\vr,\vr') \to\delta^3(\vr'-\vr)
\eeq
in the limit $\beta\to 0$.
Contour plots of $\Gdel$ and the Green function giving the 
potential, with
$\beta = 1$,
appear in Figure~1
.
Note that $\Gphi(\vr,\vr') = 
\Gphi(\vr',\vr)$, since $D^{-1}$ is self-adjoint, 
whereas $\Gvel$ and $\Gdel$ lack this symmetry.

\section{The distant survey approximation}

An interesting limit occurs for a distant-survey
(Kaiser 1987, Zaroubi \& Hoffman 1994)
in which $\vr\approx\vr'$ or
$|\vr'-\vr| \ll r$. For definiteness, 
let us choose $\vr'=\zhat$, the unit vector in the $z$-
direction.
Then 
$\dr\approx\dz$, the radial part of the Laplacian $\approx 
\dz^2$,
and the Kaiser equation reduces to 
\beq{KaiserLimitEq}
\left[\dx^2 + \dy^2 + (1+\beta)\dz^2\right]\phi = A\delta_s.
\eeq
A simple rescaling of the $z$-axis transforms the above expression into the
Poisson equation. This fact and the application of the
gradient and Laplacian operators to the Green function for the 
potential
immediately give us the three Green functions
in the distant-survey limit: 
defining 
\beq{DrDefEq}
\vdiffr
\equiv
\vr-\vr'
\equiv 
(\sin\vartheta\cos\phi,\sin\vartheta\sin\phi,\cos\vartheta)
\Delta r
\eeq
and 
$\mu_{l}\equiv\cos(\vartheta)$, with
$\vartheta$ being the angle
between $\vdiffr$ and the line-of-sight (the $z$-axis, parallel to both 
$\vr$ 
and $\vr'$ in this limit),
\beq{LimitGsolEq}
\Gphi(\vr,\vr') = 
-{1\over 4\pi\diffr}\left(1+\beta-\beta\mu_{l}^2\right)^{-1/2}\,;
\eeq
\beq{LimitGvelsolEq}
\Gvel(\vr,\vr') = 
{1\over 4\pi\diffr^2}\left(1+\beta-\beta\mu_{l}^2\right)^{-
1/2}
\left(
\ehat_r +
{\beta\mu_{l}\sqrt{1-\mu_{l}^2}\over 1+\beta-\beta\mu_{l}^2}
\ehat_\vartheta
\right)\,;
\eeq
\beq{LimitKsolEq}
\Gdel(\vr,\vr') = 
\frac{(3+\beta)\mu_{l}^2-1-\beta}
{4\pi\diffr^3\left(1+\beta-\beta\mu_{l}^2\right)^{5/2}}
+
{3+2\beta\over 3+3\beta}
\delta^3(\vr'-\vr)\,.
\eeq
Thus $\Gdel$ vanishes on the cone given by 
\beq{ConeEq}
|z'-z| = \left({1+\beta\over 2}\right)^{1/2}
\sqrt{(x'-x)^2 + (y'-y)^2},
\eeq
takes positive values inside this cone and takes negative 
values
outside of it.

\section{The real world: nonlinearity, shot noise and finite volume}

The above treatment applied to an idealized world were the Kaiser equation
was strictly valid, and where 
we could measure the redshift space density field $\ds$
accurately throughout all space. 
Alas, this is not the world in which we live. Below, we discuss the 
following three sources of error:
\begin{enumerate}
\item
The Kaiser equation, which is exact to first order, is
only accurate when $\delta\ll 1$ and thus
breaks down on small scales where nonlinear evolution
has become important.

\item
Our galaxy surveys do not measure the continuous field
$\delta_\rs$ but the discrete galaxy distribution, so our
estimates of $\ds$ are contaminated by Poissonian 
``shot noise".

\item
Our galaxy surveys sample only a finite volume, which forces us
to truncate our volume integrals and accept truncation errors in 
our reconstructed fields.

\end{enumerate}
There are of course many more potential sources of error, such as 
nonlinear biasing and flux errors to mention 
two, but we will restrict our discussion to the three 
listed problems, which are {\it always} present. 
In what follows, we also limit ourselves to reconstruction of
the density field, the analogy for the velocity field
and the peculiar gravitational potential
being obvious.

\subsection{Nonlinear clustering}

On small scales, the 
gravitational evolution has entered into the deeply non-linear regime,
where the
``fingers-of-God effect" causes virialized clusters to appear 
elongated along the line of sight.
Although some authors have attempted to extend the Kaiser equation
into the quasi-linear regime using the Zel'dovich approximation, 
an exact reconstruction of the real-space density is of course
impossible in the deeply non-linear regime. Essentially, 
this is because the gravitational $n$-body problem 
exhibits chaotic behavior, rendering the inversion 
problem numerically ill-posed.
We thus follow the same approach as all previous authors,
and smooth all our fields ($\ds$, $\deltar$, {\etc}):
\beq{SmoothingEq}
\ds(\vr) \mapsto (W\star\ds)(\vr) = \int W(\vr'-\vr)\ds(\vr') d^3r',
\eeq
the star denoting convolution.
For definiteness, let us choose a Gaussian smoothing kernel
\beq{GaussianWindowEq}
W(\vr) \equiv {1\over (2\pi)^{3/2}r_s^3} e^{-{1\over 2}(r/r_s)^2},
\eeq
and choose the {\it smoothing scale} $r_s$ to be large enough that
the resulting fields can be expected to obey linear theory\footnote{
Since smoothing suppresses high-frequency Fourier components, 
its success of course hinges on the standard assumption
that the gravitationally induced mode coupling does not 
cause a significant ``blue leak" of power from short to long
wavelengths.}.

\subsection{Shot noise}

What we measure is of course not the smooth
field $\ds$ but the discrete galaxy density 
$n(\vr)=\sum \delta(\vr-\vr_i)$, with a delta function
at the location $\vr_i$ of each galaxy.
We estimate the unsmoothed density by 
simply $\delta n/\nbar \equiv (n-\nbar)/\nbar$, 
where $\nbar$, the {\it selection function}, 
specifies the survey geometry
by giving the expected density of observed galaxies at each position in space.
Our estimate of of the redshift space density $\ds$ is thus 
the convolution of $n/\nbar - 1$ with the smoothing kernel $W$.
Hence our real-world 
generalization of \eq{KdefEq} becomes
\beq{delEstEq}
\deltar'(\vr) \equiv 
\int_V \Gdel(\vr,\vr')\left(W\star\left[{n\over\nbar}-1\right]\right)(\vr'),
\eeq
where the prime on the left hand side has been added to distinguish the estimate 
$\deltar'$ from the true value $\deltar$. We will return to the issue of how 
to choose $V$, the volume to integrate over, in the following subsection.
Since convolution is associative (simply corresponding to multiplication
in Fourier space), 
\eq{delEstEq} is equivalent to 
\beq{delEstEq2}
\deltar'(\vr) =
\int_V (W\star\Gdel)(\vr,\vr')\left[{n(\vr')\over\nbar(\vr')}-1\right],
\eeq
{\ie}, to smoothing the Green function (with respect to $\vr'$)
instead of the galaxy distribution.

\noindent
Let us, as is customary, model $n$ as a Poisson process whose intensity 
(average point density) is $\lambda(\vr)\equiv\nbar(\vr)[1+\ds(\vr)]$.
We model (the smoothed) $\ds$ as a random field 
with $\expec{\ds(\vr)}=0$ and define its correlation function as
\beq{zsCorrEq}
\xi_s(\vr,\vr') \equiv \expec{\ds(\vr)\ds(\vr')},
\eeq
but make no assumptions of it being homogeneous, isotropic or Gaussian.
Defining the {\it reconstruction error} as 
\beq{ErrorDef}
\Delta(\vr)\equiv \deltar'(\vr)-\deltar(\vr),
\eeq
we now proceed to compute the statistical properties of this error.
Since a Poisson process satisfies
\beqa{PossionEq1}
\expec{n(\vr)}&=&\lambda(\vr),\\
\label{PossionEq2}
\expec{n(\vr)n(\vr')}&=&\lambda(\vr)\lambda(\vr')+
\delta(\vr-\vr')\lambda(\vr),
\eeqa
a straightforward calculation shows that
\beqa{ErrorEq1}
\expec{\Delta(\vr)}&=&0,\\
\label{ErrorEq2}
\expec{\Delta(\vr)^2}&=& \sigma_n(\vr)^2 + \sigma_t(\vr)^2,
\eeqa
where 
\beqa{NoiseErrorEq}
\sigma_n(\vr)^2 &\equiv&
\int_V {\left[(W\star\Gdel)(\vr,\vr')\right]^2\over\nbar(\vr')}d^3r',\\
\label{TruncationErrorEq}
\sigma_t(\vr)^2&\equiv& 
=\int_{\notV}\int_{\notV}
\Gdel(\vr,\vr')\Gdel(\vr,\vr'')\xi_\zs(\vr',\vr'')
d^3r'd^3r''.
\eeqa
When computing $\sigma_n^2$, the {\it shot noise error},
the integral is to be taken over the volume $V$ (see below). 
When computing $\sigma_t^2$, the {\it truncation error},
the integrals are to be taken over all space 
{\it except} the volume $V$.
\Eq{ErrorEq1} simply tells us that
our reconstruction is {\it unbiased}, in the sense that
the expectation value of the error is zero.
The remainder of this subsection is devoted to the shot noise term in 
\Eq{ErrorEq2}.

Although the shot noise error $\sigma_n$ is easily computed numerically
using \eq{NoiseErrorEq}, it is instructive to make a crude
estimate of it. 
First, let us assume that $\nbar(\vr)$ is constant, so that we can factor it out
of the integral in \eq{NoiseErrorEq}.
Second, let us make the distant survey approximation for $\Gdel$
(this is a good approximation if $\vr$ is more than a few 
smoothing lengths away from us, as it is easy to show that
the smoothed Green function $W\star G$ falls off rapidly, 
at least as fast as $r^{-3}$, for $|\vr-\vr'|\gg r_s$, 
so that the main contribution to the integral comes from within
a few smoothing lengths of $\vr$). 
Third, 
let us overestimate $\sigma_n^2$ slightly by extending the 
integration to all space (this makes only a minor 
difference, as we will find it optimal 
to integrate considerably beyond the smoothing scale anyway). 
Now we can apply 
Parseval's theorem and the convolution theorem to the remaining 
integral, and obtain
\beq{NoiseApproxEq1}
\sigma_n^2 \approx {1\over(2\pi)^{3}\nbar}\int |\Gdelh(\vk)|^2 |\Wh(\vk)|^2 d^3k,
\eeq
where hats denote Fourier transforms. From 
\eq{KaiserLimitEq}, we immediately recover Kaiser's familiar
result
\beq{GdelhApproxEq}
\Gdelh(\vk) \approx \left[1 + \beta{\left(k_z\over k\right)^2}\right]^{-1}, 
\eeq
so substituting this into \eq{NoiseApproxEq1}, we obtain the estimate
\beq{NoiseApproxEq2}
\sigma_n^2 \approx {1\over(2\pi)^{3}\nbar}\int
{e^{-(r_s k)^2}k^2\sin\theta
\over\left[1+\beta\cos^2\theta\right]^2}
\>dk\,d\theta\,d\phi 
= \left({\arctan\sqrt{\beta}\over 8\pi^{3/2}\sqrt{\beta}}\right)
\left({1\over\nbar r_s^3}\right).
\eeq
Defining the effective {\it smoothing volume} 
\beq{SmoothingVolDefEq}
V_s\equiv \left({8\pi^{3/2}\sqrt{\beta}\over\arctan\sqrt{\beta}}\right)r_s^3,
\eeq
we can conclude this section with the familiar-looking
shot noise formula
\beq{SmoothVolEq}
\sigma_n^2 \approx {1\over\nbar V_s}.
\eeq
In other words, the crucial quantity is the expected number of 
galaxies in a smoothing volume.

\subsection{Finite volume}

Although the truncation error $\sigma_t$ is straightforward to
compute numerically given any prescribed volume $V$, 
either by evaluating the 
integral in \eq{TruncationErrorEq} or by making reconstructions from 
Monte Carlo galaxy catalogs,
we now make a very crude estimate $\sigma_t$ to obtain a 
qualitative understanding of how to best choose $V$ and the smoothing scale.
Although the unsmoothed fields $\deltar$ and $\ds$ have ample 
structure on very small scales, the corresponding 
smoothed fields are virtually featureless 
on scales $r\ll r_s$. Hence the correlation $\xi_s$ is 
almost perfect for $|\vr'-\vr|\ll r_s$, and typically 
falls off like some power law for $|\vr'-\vr|\gg r_s$.
Thus let us make the crude approximation 
\beq{xiApproxEq}
\xi_s(\vr,\vr') \approx \cases{
\sigma^2&for  $|\vr'-\vr|\ll r_s,$\cr
0&for  $|\vr'-\vr|\gg r_s$,
}
\eeq
for some constant $\sigma$ which is the {r.m.s.} fluctuation
in a smoothing volume.
The only non-negligible contributions to the integrals in
\eq{TruncationErrorEq}
thus arise when $|\vr'-\vr|\simlt r_s$. This can be interpreted as
there being a large number of independent volumes
in which the field is coherent, and that the total 
variance $\sigma_t^2$ is simply a sum of the variance arising from each 
coherence volume that we do not look at, {\ie}, that 
falls outside of $V$. 
Assuming that $V$ is large relative to the smoothing scale, 
replacing $\Gdel(\vr,\vr'')$ by 
$\Gdel(\vr,\vr')$ will thus be a good approximation, and
\eq{TruncationErrorEq}
reduces to simply 
\beq{TruncationApproxEq1}
\sigma_t(\vr)^2 \approx a\,\sigma^2 V_s 
\int_{\notV}
\Gdel(\vr,\vr')^2 d^3r',
\eeq
where $a$ is some constant of order unity.
In the distant survey approximation of
\eq{LimitKsolEq}, we see that 
$\Gdel(\vr,\vr')\propto |\vr'-\vr|^{-3}$ far away,
so if $V$ is chosen to be a sphere of radius $r_c$
around $\vr$, then 
\beq{TruncationApproxEq2}
\sigma_t^2 \simpropto \left({r_s\over r_c}\right)^3\sigma^2  
\propto {V_s\over V}\sigma^2.
\eeq

\subsection{How to choose $V$ and the smoothing scale}

In this section, we give some useful rules of thumb for how to choose
the volume $V$ and the smoothing scale $r_s$. 

Since the integrand in \eq{NoiseErrorEq} is always
positive, the shot noise error 
$\sigma_n$ always {\it increases} if we increase the
volume $V$, whereas \eq{TruncationApproxEq2} shows that 
the truncation error $\sigma_t$ {\it decreases} as we increase $V$.
Since all we care about is the total error $\sigma_n^2 + \sigma_t^2$, 
we find ourselves facing a tradeoff. 
Similarly, 
\eq{SmoothVolEq} shows that the shot noise error decreases if 
we increase the smoothing volume $V_s$, whereas 
\eq{TruncationApproxEq2} shows that the relative truncation error
$\sigma_t/\sigma$ increases  if
we increase the smoothing volume $V_s$.
Minimizing the total error
by differentiating $\sigma_n^2 + \sigma_t^2$ with respect to 
$V$ and $V_s$, we of course
end up with parameter choices such that 
the two sources of error 
are of comparable importance, {\ie}, such that 
$\sigma_n$ is of the same order of magnitude as $\sigma_t$.
Another way of phrasing this is that
given some finite level of truncation error $\sigma_t$, there is no point in 
trying to reduce the shot noise error $\sigma_n$ way below this value
(and vice versa), 
as this will cause almost no reduction in the total error.
We thus obtain one of our rules of thumb by simply equating $\sigma_n$ 
and $\sigma_t$, which after substituting 
equations\eqnum{SmoothVolEq} and\eqnum{TruncationApproxEq2}
gives
\beq{ThumbEq1}
\sigma V_s \approx \sqrt{V\over\nbar} \ ,
\eeq
so that the optimal smoothing volume
$V_s$ then goes roughly as the geometric mean of 
the total volume $V$ and the average volume per galaxy, $1/\nbar$. 

In practice, the survey specifics often place a firm lower 
limit on either $\sigma_n$ or $\sigma_t$, which makes the
best choices of $V$ and $V_s$ quite simple. We now turn to a couple of 
specific examples:

\subsubsection{Slices and pencil beams}

If the survey geometry is that of a thin slice, such as
for instance the CfA or Las Campanas redshift surveys,
or that of a so-called pencil beam,
then the dominant source of error at a point $\vr$ 
will usually be the truncation error 
$\sigma_t$ caused
by the most nearby edge of the slice/pencil.
This means that we should choose the 
smoothing scale as small as we feel comfortable with given the
non-linearities on small scales, perhaps around 
$r_s\approx 10 h^{-1}\Mpc$. 
\Eq{TruncationApproxEq1} shows that
since $\Gdel^2$ falls off so rapidly (about as $|\vr'-\vr|^{-6}$), 
the truncation 
error will be completely dominated by the smallest distance from $\vr$ to 
the edge of the volume $V$.  In other words, given any volume $V$, 
we can get almost as small truncation errors by replacing it by the
largest sphere around $\vr$ that will fit inside it,
and this defines a characteristic $\sigma_t$ that we refer to below.
In practice, the reconstruction calculations are so 
fast that there is little need for such attempts to save computer time,
especially considering the amount of effort that has gone into the 
preceding data collection. 
Rather, the main concern is that we want to avoid extending 
$V$ out to such large radii that the shot noise error 
becomes as large as $\sigma_t$. Since $\nbar(\vr)$ is 
usually a decreasing function of $r$, \Eq{SmoothVolEq} shows that this
would happen if $r$ where allowed to be so large that
$\nbar(r) \simlt 1/\sigma_t V_s$.
In conclusion, when reconstructing the real-space fields at a point 
$\vr$, we will do fine if we simply choose $V$ to be 
the entire slice volume, but truncated at 
some cutoff radius $r_c$ such that 
\begin{itemize}
\item $r_c - r$ exceeds the sideways 
distance from $\vr$ to the edge of the slice/pencil beam
\item $\nbar(r_c) \simgt 1/\sigma_t V_s$.
\end{itemize}

\subsubsection{``Non-skinny" surveys}

For surveys where all three dimensions are large relative to the 
smoothing length (such as the IRAS 1.2Jy all-sky survey),
the situation is quite different.
When reconstructing the real-space field at 
points $r$ that are
many smoothing lengths away from the survey boundaries, 
truncation errors will not a priori dominate as above.
Hence an efficient analysis strategy is to simply
truncate the data at some radius $r_c$ where $\nbar$
begins to fall off rapidly, take $V$ to be the the entire remaining 
survey volume (since $\sigma_n$ saturates a few
smoothing lengths away from $\vr$ anyway), 
and then select the smoothing scale according 
to \eq{ThumbEq1}.

\section{Examples}

\label{tee_bone}

We now evaluate the Green function method in 
two simple numerical examples. First, for a 
qualitative assessment, we realize a single 3D field in real space,
remap the density into redshift space with peculiar flows derived from
linear theory,
and then perform the reconstruction of the real density field with
the Green function method.
The results of this procedure are given 
in Figure~2 for an $\Omega = \beta = 1$ 
Universe, 
where 2D slices of the density fields are shown. 
In addition to the reconstructed field (lower-left plot) we
provide a reconstruction using an incorrect $\beta$ 
value of 0.25 (lower-right plot).
This illustrates the sensitivity of the method to $\beta$, a feature
which may be useful in constraining $\Omega$ and the degree of linear bias.

Details of the procedure that led to the data in Figure~2
are as follows: We start with a triply periodic
250~$h^{-1}$~Mpc cube and
realizes a Gaussian random field on grid of $64^3$ points.
The power spectrum of density fluctuations has the
form of standard cold dark matter (e.g., Efstathiou, Bond \& White 1992),
smoothed with a Gaussian filter on scales
of 50~$h^{-1}$~Mpc. The normalization is chosen near the 
COBE value to approximate
observed large-scale structure; 
before smoothing, the r.m.s. density
fluctuations in a sphere of radius 8~$h^{-1}$~Mpc are at a level of 1.1 
relative to the mean density, \ie, $\sigma_8=1.1$.
We then estimate the velocity field at the grid points and displace the
the points by an amount appropriate for an observer at some chosen origin to
mimic the distribution in redshift space. 
These points are interpolated onto a lower 
resolution grid ($32^3$) to reduce discreteness effects.
This lower-resolution grid represents the volume-limited
red-shift space field that is used in the reconstruction.
For the 2D plots in Figure~2,
we perform the reconstruction on a plane of grid points; 
at each point a 3D integral over the redshift-space grid 
is evaluated with the kernel of \eq{KsolEq}
under the assumption that the density field is a set of 
weighted Dirac-delta functions on the grid. 
For speed enhancement, the sum in \eq{KsolEq} was evaluated
at a grid of points $(x,\mu)$ once and for all in 
a lookup table.
When $\vr\approx\vr'$, the order zero analytic reconstruction
kernel (Ozark) of \eq{LimitKsolEq} was used.

Evidently, the reconstruction works well for 
regions away from the edges of the mock survey in Figure~2.
The reconstruction has recovered both
the amplitude of the original field and the topology of the
isodensity contours.
However, a qualitative assessment of the errors is in order.
Our focus in this section is on truncation errors as well as 
errors incurred during interpolation
and integration. 
We estimate the errors with a Monte Carlo method;
we realize 72 random density fields, perform reconstruction
on each as described above, and determine the r.m.s. point-wise
difference between the
real density field and the reconstruction. 
Figure~3 shows the r.m.s.\ error in reconstruction for points on one of the
principle axes of the cubical survey volume (the points lie on a path
which bisects the plane shown in Figure~2).

Truncation effects are clear in Figure~3 
from the increase in error near the boundary
of the survey volume. The truncation error worsens as the
correlation length of the density field increases, as can be seen
from a comparison of the error curves with correlation lengths
of 50 and 100~$h^{-1}$~Mpc (induced
by smoothing on those scales);
large correlation lengths correspond to low frequency modes which are 
more susceptible to the ill effects of a finite survey volume.

For the case of the realizations with a smoothing scale of 50~$h^{-1}$~Mpc,
we estimate the theoretical truncation error by directly 
integrating \eq{TruncationErrorEq}.
The relatively high dimensionality of
the integral and the rapid variation of the integrand near the
boundaries of the survey volume suggest that this problem is best handled
by parallel computation (we are grateful for time on a Cray T-3D
at Jet Propulsion Laboratory). 
For reference, an approximate theoretical error curve is shown in Figure~3. 
The remaining error comes from grid interpolation,
the numerical integrator, and to a lesser extent the breakdown
in the Kaiser equation, which is of course only exact for 
$\delta\ll 1$. This latter source of error scales roughly as
$\delta^2$, and does become more important
at smaller smoothing scales.
While in this example
we are content to let these errors lie between 5\% and 10\% near the
center of the survey,
they may be further reduced with a finer grid mesh and a higher-order
integrator.

\section{Discussion}

We have presented the Green functions for computing 
the
gravitational potential, the peculiar velocity field and the 
real-space
density fluctuations from the distribution of matter in 
redshift space.
Our results are based on Kaiser's (1987) analysis of 
redshift-space distortions in the linear regime, 
and are applicable to surveys of arbitrary
geometry.

One virtue of the method presented here is that the entire 
calculation is performed in real space. 
This means that the effects of 
complications such as survey boundaries and
the presence of galaxy clusters,
which are manifestly in physical space rather than Fourier space,
are easy to understand and quantify.
For example, the magnitude
of the Green function in the parts of
space that we are ignoring immediately gives an indication of
how much information we are loosing, {\ie}, on the magnitude of the
error bars on the reconstructed field.
As we have discussed, this allows 
the problems caused by shot noise, 
survey boundaries and nonlinear clustering
to be minimized by appropriate choices
of the smoothing scale and the integration volume $V$.

We wish to emphasize several possible uses of the Green 
function method.
The first is toward 
ferreting out estimates of 
the linear growth parameter 
$\beta = \Omega^{0.6}/b$.
While $\beta$ has been estimated in elegant manners from the
correlation anisotropy (e.g., Hamilton 1992),
it may also be determined with the Green function $\Gdel$:
The best estimate for $\beta$ is the value which minimizes 
the anisotropy in the real-space two-point correlation function
of the reconstructed field.
Quantitative conditions for minimizing the anisotropy in 
$\xi_\rs$ may be
derived with an orthogonal decomposition in spherical 
harmonics 
(e.g., Hamilton 1992). For example, a least-squares criterion 
may be imposed on the amplitudes of terms with
spherical harmonics $Y_{lm}$ for which $l > 0$, with greatest 
weights
assigned to 
quadrupole terms, as they couple strongly to the redshift-space 
distortions.

The second application of the Green function method is to 
extract
information on the nature of biasing. 
Specifically, the reconstructed real-space density field
contains the implicit assumption of a linear bias between 
luminous and dark matter.
Thus it is worth comparing the field with density
reconstructions from other techniques such as the
POTENT algorithm (Bertschinger \& Dekel 1989). While such a 
comparison
will not alone break the degeneracy between $b$ and $\Omega$, 
it can
test the validity of the linear biasing hypothesis.

Ultimately, the method's most important use is the estimation 
of
peculiar velocities. Even linear peculiar flows can provide 
stringent 
constraints on cosmological models (e.g., Lauer \& Postman 
1994;
Tegmark, Bunn \& Hu 1994).
Since the Green function method generates peculiar
velocities directly from redshifts without the difficult 
measurement
of real-space distances, it may have fruitful application in 
the
large-scale digital sky surveys 
that are currently coming on line.

Our purpose here has been to provide the mathematical
underpinnings of the Green function approach to cosmic field 
reconstruction from redshift data. 
We are now embarking on the next step, the analysis of redshift surveys.

\mbox{}

We thank James Anglin, Marc Davis, Andrew Hamilton, Saleem Zaroubi
and our referee for useful comments and suggestions.
BCB acknowledges support from the NASA HPCC program
and help from Isabel Dulfano. 

\newpage

\begin{center}
{\sc References}
\end{center}

\begin{list}{}%
{
  \setlength{\leftmargin}{0.5in}
  \setlength{\rightmargin}{0.0in}
  \setlength{\itemindent}{-0.5in}
  \setlength{\itemsep}{0.0in}
}

\item Bertschinger, E., \& Dekel, A. 1989, ApJ, 336, L5

\item Bromley, B.\ C. 1994, ApJ, 423, L81

\item Cole, S., Fisher, K.\ B., \& Weinberg, D. \ H. 1994,
MNRAS, 267, 785 

\item Efstathiou, G., Bond, J.\ R., \& White, S.\ D.\ M. 1992, MNRAS, 258, 1P

\item Fisher, K.\ B., Scharf, C.\ A., \& Lahav, O. 1994, 
MNRAS, 266, 219

\item Giovanelli, R., \& Haynes, M.\ P. 1991, Ann. Rev. Astr. 
Ap., 29, 499

\item Gramann, M., Cen, R.\ Y. \& Gott III, J.\ R. 1994, ApJ, 425, 382

\item Hamilton, A.\ J.\ S. 1992, ApJ, 385, L5

\item Hamilton, A.\ J.\ S. 1993, ApJ, 406, L47

\item Kaiser, N. 1987, MNRAS, 227, 1

\item Lauer, T. \& Postman, M. 1994, ApJ, 425, 418

\item Nusser, A. \& Davis, M. 1994, ApJ, 421, L1


\item Peebles, P.\ J.\ E., 1993, Principles of Physical 
Cosmology
(Princeton: Princeton University Press)

\item Smoot, G.\ F. et.al. 1992, ApJ, 396, L1


\item Taylor, A. \& Rowan-Robinson, M. 1993; MNRAS, 265, 809

\item Tegmark, M., Bunn, E., \& Hu, W. 1994, ApJ, 434, 1

\item Yahil, A., Strauss, M.\ A., Davis, M., \& Huchra, J.\ 
P. 1991, ApJ, 372, 380

\item Zaroubi, S. \& Hoffman, Y. 1994, preprint

\end{list}

\newpage

\begin{figure}[phbt]
\centerline{\epsfxsize=5.0in\epsfbox{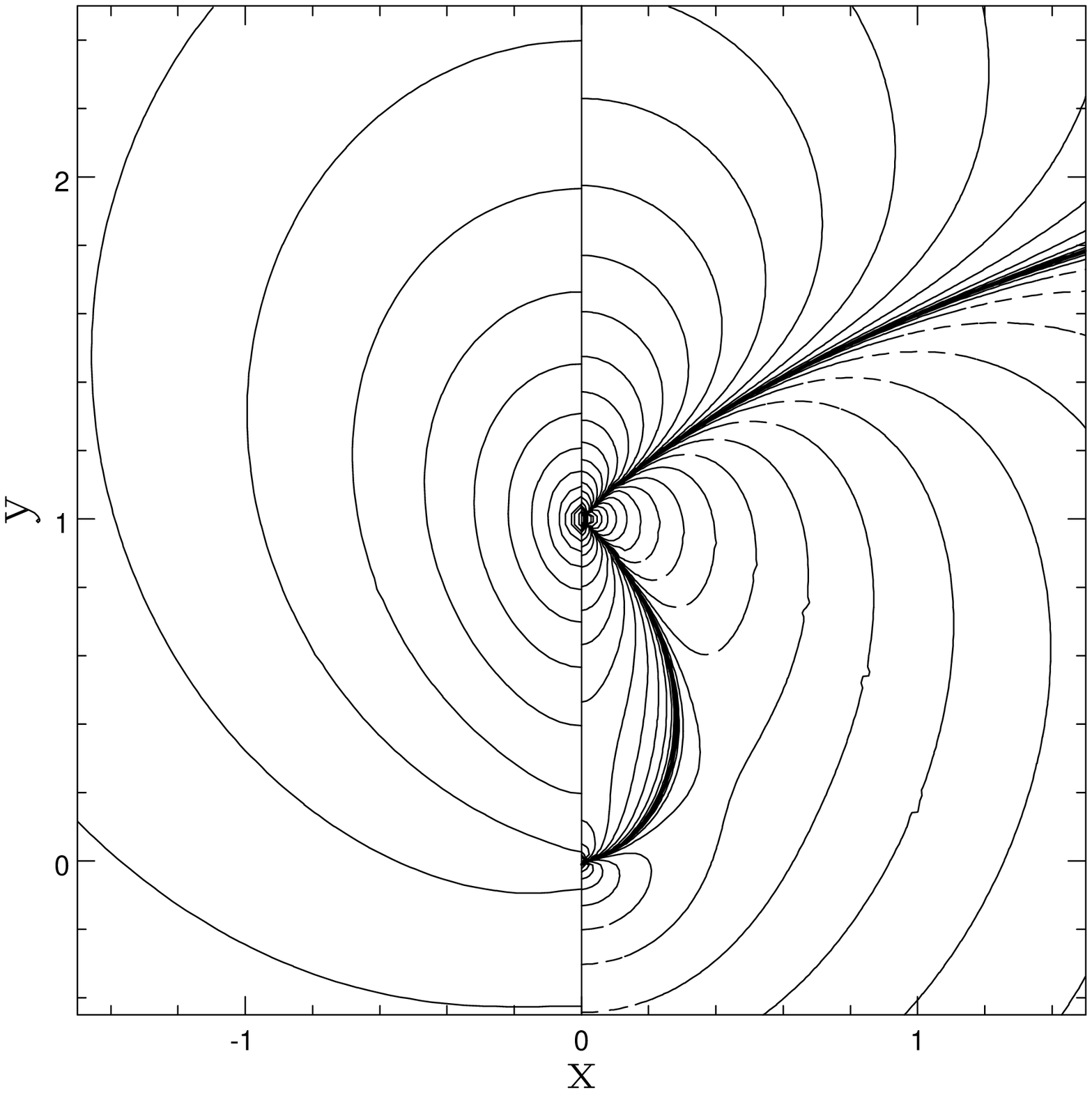}}
\caption{
The Green functions
$\Gphi$ and $\Gdel$, for $\beta=1$.
}

To the left of the $y$-axis are contours of 
$\Gphi(\vr,\vr')$, 
while the contours 
to the
right correspond to 
$\Gdel(\vr,\vr')$ for
$\vr = (x, y, 0)$ and $\vr' = (0,1,0)$.
The light solid lines
and dashed contours 
indicate positive and negative values, respectively, 
with logarithmic spacing. 
The heavy solid line is the $\Gdel(\vr) = 0$ contour.
The trend in both functions is toward
high absolute values near the center of the figure.
Values of these functions for more general choices of
$\vr$ and $\vr'$ follow directly from symmetry and scaling properties.
The functions essentially give the contribution of a point mass
at $\vr$ to the reconstructed field at $\vr'$ 
(point at center of the figure).
\label{fig:green}
\end{figure}

\newpage

\begin{figure}[phbt]
\centerline{\epsfxsize=5.0in\epsfbox{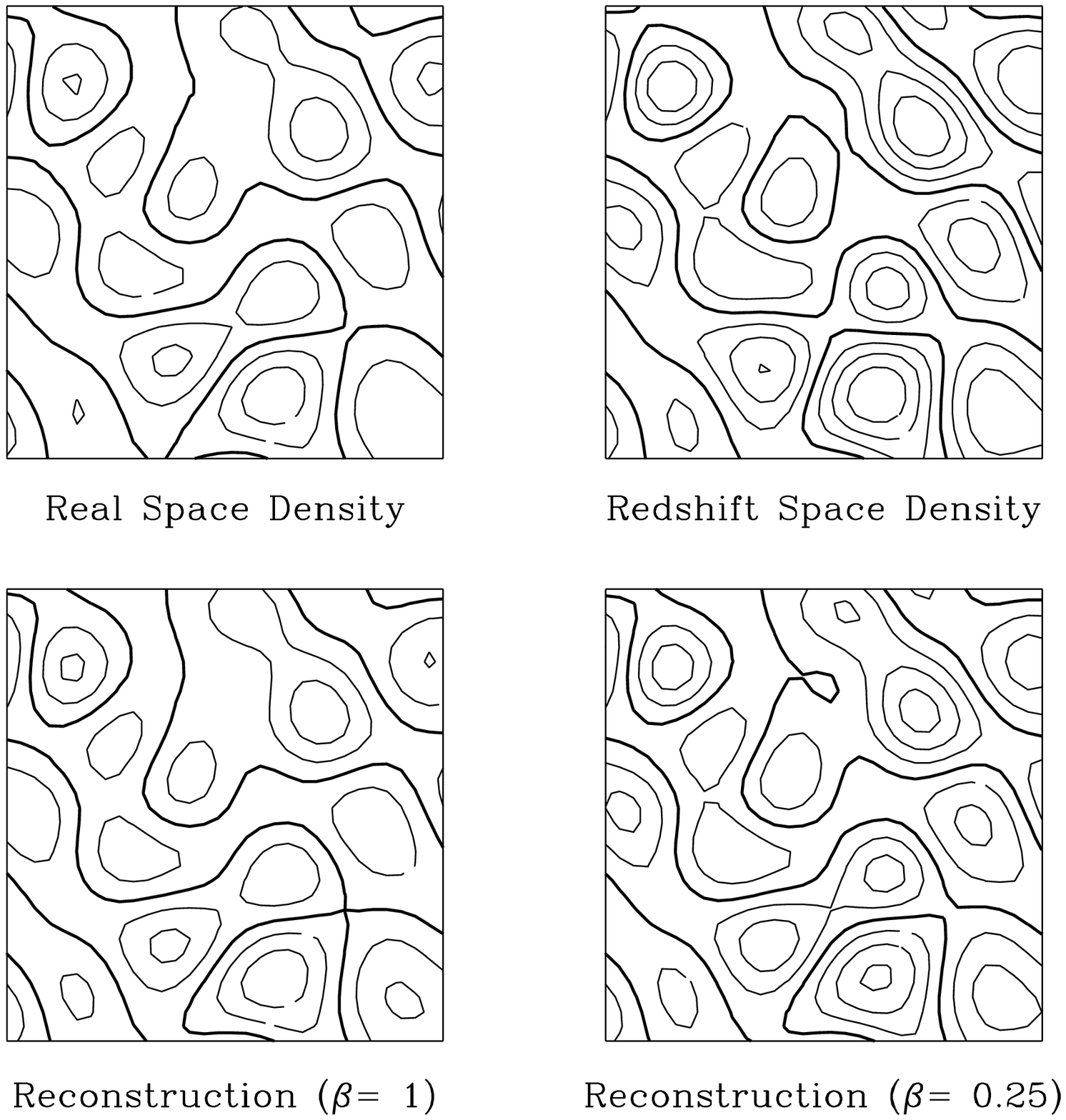}}
\caption{ An example of the reconstruction method.} 
As discussed in the text, the reconstruction method is applied to a
linearly evolved Gaussian random field in a 250~$h^{-1}$~Mpc cube.
The linearly spaced contours, denoting isodensity surfaces at a 
mid-plane of the cube,
show overdensities (solid lines), underdensities (broken
lines), 
and the mean density (heavy solid lines).
The observer is located at a corner of the cube.
The upper-left plot is the density in real space, generated
from a CDM power spectrum with  50~$h^{-1}$~Mpc smoothing. 
The corresponding redshift-space density (upper-right) is used to 
reconstruct the real-space fields; the lower-left plot was 
derived with
the correct $\beta$ value while an incorrect value yielded 
the plot on the
lower right. 

\label{fig:recon}
\end{figure}

\newpage

\begin{figure}[phbt]
\centerline{\epsfxsize=5.0in\epsfbox{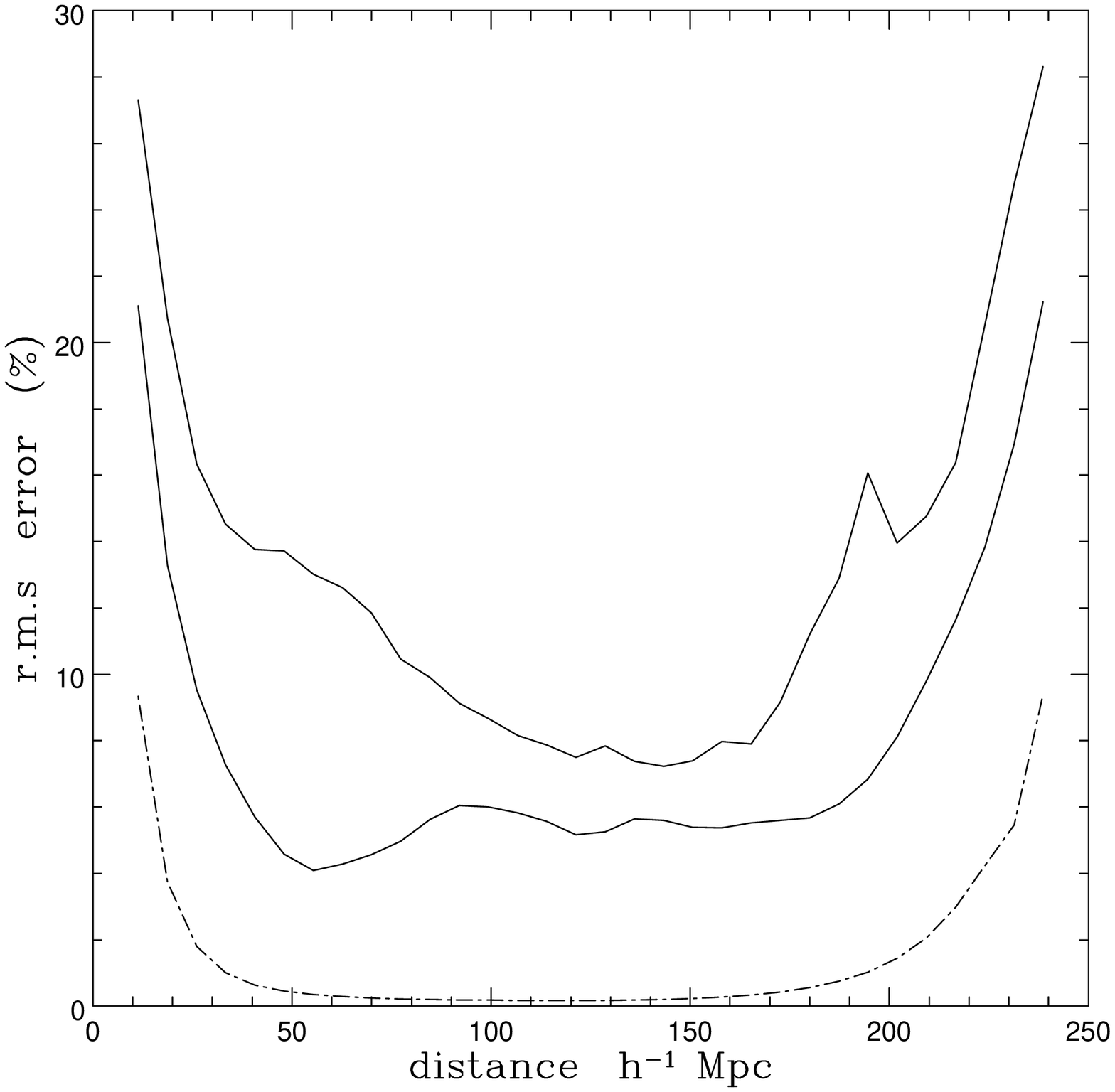}}
\caption{Density reconstruction errors.}
The r.m.s.\ reconstruction errors, $\sigma(\vr)/\sqrt{\xi_r(0)}$, 
are estimated from 72 realizations of a
linearly evolved CDM field, with 100~$h^{-1}$~Mpc smoothing (top curve)
and 50~$h^{-1}$~Mpc smoothing (middle curve), in the survey region
used for Figure~2.
The horizontal axis is distance along a principle axis of the cube.
The lower (dashed) curve
is an estimate of the expected truncation noise, $\sigma_t(\vr)$ for
the case of 50~$h^{-1}$~Mpc smoothing.
\label{fig:recerr}
\end{figure}

\end{document}